\documentclass{bmcart}

\usepackage[utf8]{inputenc} 

\usepackage{amsmath}
\usepackage{multirow}
\usepackage{graphicx}
\usepackage{caption,subcaption}

\startlocaldefs
\endlocaldefs

\begin{document}
	
	\begin{frontmatter}
		
		\begin{fmbox}
			\dochead{Methodology}
			
			
			\title{Spatio-Temporal Multi-step Prediction of Influenza Outbreaks}
			
			
			\author[
			addressref={aff1},           
			email={jie-zhang@g.ecc.u-tokyo.ac.jp}   
			]{\inits{JZ}\fnm{Jie} \snm{Zhang}}
			\author[
			addressref={aff1},
			email={nawata@tmi.t.u-tokyo.ac.jp}
			]{\inits{KN}\fnm{Kazumitsu} \snm{Nawata}}			
			\author[
			addressref={aff2},
			corref={aff2},                       
			email={hy.wu@siat.ac.cn}
			]{\inits{HW}\fnm{Hongyan} \snm{Wu}}
			
			\address[id=aff1]{
				\orgname{Department of Technology Management for Innovation, The University of Tokyo}, 
				\street{Hongo},                     %
				\postcode{1138656}                                
				\city{Tokyo},                              
				\cny{Japan}                                    
			}
			\address[id=aff2]{
				\orgname{Joint Engineering Research Center for Health Big Data Intelligent Analysis Technology, Shenzhen Institutes of Advanced Technology, Chinese Academy of Sciences}, 
				\street{1068 Xueyuan Avenue, Shenzhen University Town},                     %
				\postcode{518055}                                
				\city{Shenzhen},                              
				\cny{China}                                    
			}
			
			
			\begin{artnotes}
			\end{artnotes}
			
		\end{fmbox}
		
		
		\begin{abstractbox}
			
			\begin{abstract} 
				\parttitle{Background} 
				Flu circulates  all over the world. The worldwide infection places a substantial burden on people’s health every year.  Regardless of the characteristic of the worldwide circulation of flu, most previous studies  focused on regional prediction of flu outbreaks. The methodology of considering the spatio-temporal correlation could help forecast  flu outbreaks more precisely. Furthermore, forecasting a long-term flu outbreak, and understanding flu infection trend more accurately could help hospitals, clinics, and pharmaceutical companies to better prepare for annual flu outbreaks. Predicting a sequence of values in future, namely, the multi-step predication of flu outbreaks should cause concern. Therefore, we highlight the importance of developing spatio-temporal methodologies to perform multi-step prediction of worldwide flu outbreaks.
				
				\parttitle{Results} 
				We compared the MAPEs of SVM, RF, LSTM models of predicting flu data of the 1-4 week(s) ahead with and without other countries' flu data. We found the LSTM models achieved the lowest MAPEs in most cases. As for countries in Southern hemisphere, the MAPEs of predicting flu data with other countries are higher than those of predicting without other countries. For countries in Northern hemisphere, the MAPEs of predicting flu data of the 2-4 weeks ahead with other countries are lower than those of predicting without other countries; and the MAPEs of predicting flu data of the 1-weeks ahead with other countries are higher than those of predicting without other countries, except for UK.

				\parttitle{Conclusions} 
				In this study, we performed the spatio-temporal multi-step prediction of influenza outbreaks. The methodology considering the spatio-temporal features  improves the multi-step prediction of flu outbreaks. 
				
			\end{abstract}
			
			
			\begin{keyword}
				\kwd{sample}
				\kwd{article}
				\kwd{author}
			\end{keyword}
			
			
		\end{abstractbox}
		%
		
	\end{frontmatter}
	
	
	
	\section*{Background}
	Influenza, short for flu, is an acute respiratory infection caused by flu viruses. Flu circulates in all over the world. The worldwide infection places a substantial burden on people's health every year. According to World Health Organization(WHO)’s report, flu is estimated to result in about 3 to 5 million cases of severe illness, and about 290 000 to 650 000 deaths. Accurately forcasting of influenza outbreaks could help taking appropriate actions, such as school closure, to prevent or reduce flu illness.
	
	Regardless of the characteristic of its worldwide circulation of flu,  most previous studies have focused on regional prediction of flu outbreaks \cite{wang2016regional,kane2014comparison,malik2017distressed,wu2017time} for two probable reasons. First, different locations in one country or one region, to some extent, share similar geo-locational characteristics, such as humidity and temperature. Flu virus shows a sensitivity to temperature and humidity.  As a result, predicting flu outbreaks of one country or one region is considered reasonable and approachable. Second, flu virus transmission is believed to occur mostly over relatively short distances. Usually, flu virus is spread through the air from coughs or sneezes. When an infected person coughs or sneezes, droplets containing viruses (infectious droplets) are dispersed into the air and can spread up to one meter, and infect persons in close proximity who breathe these droplets in.
	
	However, one fast-growing risk group, travelers, is  neglected from these overviews. Several changes in our globalizing world contribute to the growing influence of the traveller group: (i) steady increase in total travel volume worldwide, (ii) advent of mass-tourism and (iii) increasing numbers of immune-compromised and elderly travelers. International sporting events and festivals as well as  traveling by airplane or cruise ship could facilitate flu virus transmissionand therefore global spread of influenza \cite{goeijenbier2017travellers}. The study in \cite{he2015global} shows that flu outbreaks correlate with each other in all countries around the world. The methodology of considering the correlation could help  forecast the flu outbreaks.
	
	Furthermore, forecasting a longer-term flu outbreak, and knowing its outbreak trend more accurately could help hospitals, clinics, and pharmaceutical companies to better prepare for annual flu outbreaks. First, manufacturing flu vaccine is a challenging work. According to WHO’s report, vaccination is the most effective way to prevent the disease. During 2015-2016 flu seasons, flu vaccine prevented an estimated 5.1 million illnesses, 2.5 million medical visits, 71,000 hospitalizations, and 3,000 pneumonia \& influenza (P\&I) deaths. The problem is that flu virus undergoes high mutation rates and frequent genetic re-assortment (combination and rearrangement of genetic material). This characteristic of flu complicates the procedure of flu vaccines production. In Februaries, World Health Organization (WHO) assesses the strains of flu virus that are most likely to be circulating over the following winter. Then, vaccine manufacturers produce flu vaccines in a very limited time. Usually, the first batch of vaccine is unavailable until September. As a result,  in an extremely limited time, manufacturers have to prepare enough vaccines \cite{gerdil2003annual,lubeck1980antigenic}. Second, beds assignment to flu patients is another challenging task due to the limited capacity of hospital beds, time-dependencies of bed request arrivals, and unique treatment requirements of flu patients. Flu seasons vary in timing, severity, and duration from one season to another. Therefore, flu hospitalization also varies greatly by sites and time in each season \cite{puig2014first}. Predicting a sequence of values in future,namely,the multi-step predication of flu outbreaks should cause concern.

	Therefore, we highlight the importance of developing global methodologies to perform multi-step prediction of worldwide influenza outbreaks. Nonetheless, not many past studies focused on multistep prediction of influenza outbreaks. The probable reason could be that multistep prediction usually results in poor accuracy due to some insuperable problems, such as error accumulation, etc. \cite{zhang2013iterated,akhlaghi2017adaptive}. One compromising method is that one can aggregate raw data to a larger time unit and then use the single-step prediction to replace multi-step prediction. For instance, if raw data is weekly based, we can aggregate weekly values to monthly values and then perform single-step prediction of the total value of the coming month (roughly around four weeks). However the aggregation hinders us from understanding the internal variation during the coming four weeks. 
	
	In this study, we performed multi-step prediction by leveraging Long Short Term Memory (LSTM). The LSTM is a special kind of RNN. In theory, the complex structure (layers and gated cells) enables LSTM to learn long-term dependencies \cite{hochreiter1997long}, simulate nonlinear function, and refine time-series prediction very well \cite{gers1999learning}.

	\section*{Methods}
	
	As shown in the Figure 1, to perform spatio-temporal flu prediction based on historical data, firstly, we scraped flu data of all the 155 countries from the FluNet, a global web-based tool for flu virological surveillance in WHO. We selected 23 countries as features since other countries have N/As in their flu data. We selected spatio-temporal related features. Then, we send those features into a model combined with LSTM and fully connected layers. Finally, the model predicts the flu data of the 1-, 2-, 3-, and 4-week ahead with other countries' flu data. To compare the results, we also predicts the flu data of the 1-, 2-, 3-, and 4-week ahead without other countries' flu data. The following subsections presents the details.

	\subsection*{ Data acquisition}
	FluNet is a global web-based tool for flu virological surveillance \cite{fluNet}. The data at country level are available and updated weekly. From FLuNet, we collected the flu data of 155 countries around the world from the 1st week of 2010 to the 18th week of 2018. We select 23 countries, the flu data of which have no NAs. The 23 countries are Australia, Brazil, Cambodia , China , Egypt , French Guiana , Ghana , Indonesia , Iran , Iraq , Ireland , Japan , Netherlands , Nicaragua, Niger, Norway, Panama, Poland, Republic of Korea, Russia, United Kingdom of Great Britain and Northern Ireland (UK), United States of America (USA).
	
	\subsection*{ Feature Selection}
	The features are selected or generated considering the spatital and temporal influence of flu outbreaks.

	\subsubsection*{Temporal factors}
	
	In temperate climates, flu outbreaks occur mainly during winter; while in tropical regions, flu outbreaks occur throughout the year. Considering the possible one-year long period of flu outbreaks our  previous studies compared the performance of the time lags of 2, 4, 9, 13, 26, and 52 week, and found that 52 weeks lead to the best accuracy \cite{zhang2017comparative}. The temperature changes could affect flu virus, and people tend to get illness. Therefore  we construct the temporal factors with three kinds of data: the original data of the past 52 weeks; the first order difference; the mean, median, standard deviations(std), maximum, and minimum of windows, the length of which are 1, 2, 3, 4, 9, 13, 26, 52 weeks.
	
	\subsubsection*{Spatial factors}
	
	Considering the global spread of influenza and the correlation between countries, we use the historical flu data of another above-mentioned 22 countries  as the prediction  features when predicating one country. Therefore  when we are predicating the flu outbreaks of one country, the other countries could affect the outcome by adjusting their weight parameters.  By this way, we get another 1,144 (22 times 52) features.
	
	\subsection*{Multi-step Prediction}
	
	There are two types of prediction of flu outbreaks: (a) single-step prediction: predicting the coming value in future by analyzing observed values in the past; and (b) multistep prediction: predicting a sequence of values in future by analyzing observed values in the past.  The idea (a) tend to accumulate the errors induced in the previous steps to future predictions. In this study we use multiple single-output prediction (MSOP) to implement multi-step prediction.
	MSOP predicts the coming several values by the same past values. In other words, when predicting $X_{t+p(p>=2)}$, MSOP jumps $X_{t+p-1 (p>=2)}$,$X_{t+p-2 (p>=2)}$, …, and $X_{t+1}$. Formula 3 explains the algorithm of MSOP. Its flow are presented in Figure 2.
	
	\begin{equation}
	\begin{split}
	X_{t+1}(predicted)& =LSTM\_MODEL\_\#01[X_{t}(observed), X_{t-1}(observed), …, X_{t-51}(observed)]\\
	X_{t+2}(predicted)& =LSTM\_MODEL\_\#02[X_{t}(observed), X_{t-1}(observed), …, X_{t-51}(observed)]\\
	X_{t+3}(predicted)& =LSTM\_MODEL\_\#03[X_{t}(observed), X_{t-1}(observed), …, X_{t-51}(observed)]\\
	X_{t+4}(predicted)& =LSTM\_MODEL\_\#04[X_{t}(observed), X_{t-1}(observed), …, X_{t-51}(observed)]\\
	\end{split}
	\end{equation}
	
	As shown in Figure 2, to predict $x_{t+1}$, we train a model by using $X_{t}, X_{t-1}, X_{t-2}, …, X_{t-52}$ as features. To predict $X_{t+2}$, we train another model by  still using $X_{t}, X_{t-1}, X_{t-2}, …, X_{t-52}$ as features. Although we use the same feature space in these two models, the two models  are trained differently with different responses ($x_{t+1} and x_{t+2}$). The research in \cite{zhang2017comparative} shows the 3-layered LSTM is efficient enough in predicting flu outbreaks.

	\subsection*{ Metrics}
	
	Because the population of some countries  is quite small and only 1 or 2 flu patients every week are reported, the study on those countries is insignificant. 
	We predicted the flu data of the coming weeks in countries with a large population. We selected Australasia, Brazil, China,  Japan, UK, and USA when considering  population and location.  
	
	We investigated the distribution of the flu data, and found that it was non-normal distribution. In our opinion, comparing models’ accuracy by Mean Absolute Percentage Error (MAPE, as shown in the Formula \ref{mape}) reflects the difference based on the median, while comparing models’ accuracy by Root Mean Square Error (RMSE) is based on means. Therefore, we used MAPE as a metrics to compare predicting the accuracy of models.
	
	\begin{equation}
	MAPE=\frac{1}{n}\sum _{t=1}^{{n}_{x}}\left|\frac{F_{t}-A_{t}}{A_{t-1}}\right|\label{mape}
	\end{equation}

	\section*{Results}
	
	Table 1 presents the MAPEs of RF, SVM, and LSTM models with and without other countries’ flu data. For example, when forecasting China’s flu data of 1-, 2-, 3-, and 4-week ahead, the MAPEs of the LSTM models with other countries’ flu data are 13.1\%, 19.8\%, 26.7\%, 36.2\%; while the MAPEs of the LSTM models without other countries’ flu data are 12.5\%, 20.2\%, 29.0\%, and 36.7\%.
	
	Figure 3 compares the MAPEs of SVM, RF, LSTM models of predicting flu data of the 1-, 2-, 3-, and 4-week ahead with other countries' flu data. In most cases, the LSTM models achieved the lowest MAPEs.
	
	Alike, Figure 4 compares the MAPEs of SVM, RF, LSTM models of predicting flu data of the 1-, 2-, 3-, and 4-week ahead without other countries' flu data. In most cases, the LSTM models achieved the lowest MAPEs.
	
	Figure 5 compares the MAPEs of the LSTM models with and without other countries’ flu data. As for countries in Southern hemisphere, i.e. Australia and Brazil, the MAPEs of predicting flu data of the 1-, 2-, 3-, and 4-week ahead with other countries are higher than those of predicting without other countries. For countries in Northern hemisphere, i.e. China, Japan, UK, and USA, the MAPEs of predicting flu data of the 2-4 weeks ahead with other countries are lower than those of predicting without other countries. Interestingly, when predicting flu data of the 1 week ahead, the MAPEs of predicting with other countries are usually higher than those of predicting without other countries, except for UK.

	\section*{Discussion}
	
	We found, in southern hemisphere (Australia and Brazil), the MAPEs of predicting flu data with other countries are higher than the MAPEs without other countries. The probable reasons are the southern hemisphere’s countries have totally different flu seasons since their winters are in June, July and August. And the countries selected in this study are mostly in Northern hemisphere and their flu data are barely correlated to the flu data of southern hemisphere’s countries. In addition, Australia is geographically isolated from other countries.  As for Northern hemisphere, the MAPEs of predicting flu data of the 2-4 weeks ahead with other countries are lower than those without other countries. That is because of high correlations among flu data of Northern hemisphere’s countries. However, the MAPEs of predicting flu data of the 1 week ahead with other countries are lower than those without other countries. That is probably because flu infection in other countries does not impact the of flu infection in target countries in one week ahead because of geographical distance. 
	
	The best MAPEs of LSTM models achieved were still very high because we used the flu data in 2017-2018 as a testing set. The 2017-2018 flu season, a pandemic-alike season, is quite different from and seriously heavier the past few seasons. And using other machine learning metrologies, such as SVR and RF,  result in higher MAPEs. In this study, we used only historical values. To some extent, historical values are a reflection of all possible related factors. However, one might say other features, such as temperature and humidity, could help predict more accurately, especially at turning points. For one thing, when we predict future values, we have to use predicted data, e.g. weather forecast. The predicting error of predicted data could intensively enlarge the predicting error in further steps. For another, how to express one country’s weather could be another problem if the country has a large area and population. A possible solution could be using two convolutional neural networks to extract features of weather and population of the whole country.
	
	\section*{Conclusions}
	In this study, we performed the spatio-temporal multi-step prediction of influenza outbreaks.The methodology considering the spatio-temporal features  improves the multi-step prediction of flu outbreak. We compared the MAPEs of SVM, RF, LSTM models of predicting flu data of the 1-4 week(s) ahead with and without other countries' flu data. We found the LSTM models achieved the lowest MAPEs in most cases. As for countries in Southern hemisphere, the MAPEs of predicting flu data with other countries are higher than those of predicting without other countries. For countries in Northern hemisphere, the MAPEs of predicting flu data of the 2-4 weeks ahead with other countries are lower than those of predicting without other countries; and the MAPEs of predicting flu data of the 1-weeks ahead with other countries are higher than those of predicting without other countries, except for UK. 
	
	\begin{backmatter}
		
		\section*{Competing interests}
		The authors declare that they have no competing interests.

		
		\bibliographystyle{bmc-mathphys} 
		\bibliography{bmc_article}      
		
		
		
		
		\section*{Figures}
		\begin{figure}[h!]
			\caption*{\csentence{The flow chart of our method.}
				The flow of our method is composed of four parts:acquire data;Select countries;features; and build models.}
		\end{figure}
		
		\begin{figure}[h!]
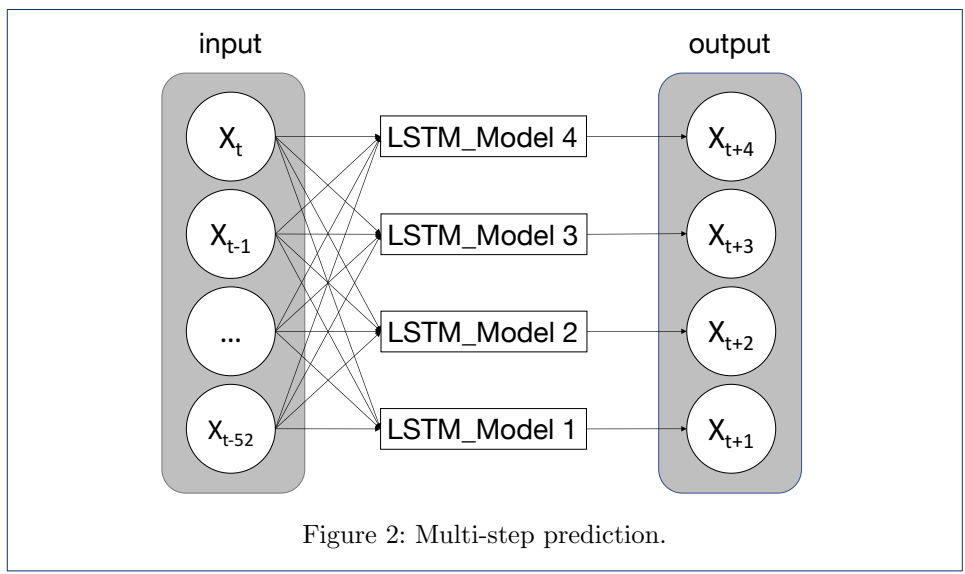

			\caption*{\csentence{Multi-step prediction.}
			}
		\end{figure}
		
		\begin{figure}[h!]
			\caption*{\csentence{The MAPES considering spatio-temporal features.}
				The figure  illustrates the MAPEs of predicting flu data of the 1-, 2-, 3-, and 4-week ahead with other countries' Flu data  .}
		\end{figure}
		
		\begin{figure}[h!]
			\caption*{\csentence{The MAPES without considering spatio-temporal features.}
				The	figure  illustrates the MAPEs of predicting flu data of the 1-, 2-, 3-, and 4-week ahead without other countries' Flu data .}
		\end{figure}
		
		\begin{figure}[h!]
			\caption*{\csentence{The result of predicting flu outbreaks using LSTM.}
				The figure  illustrates the comparison of MAPEs of predicting flu data of the 1-, 2-, 3-, and 4-week ahead with other countries' flu data and those without other countries' flu data}
		\end{figure}
		
		
		\section*{Tables}

		\begin{table}[h!]
			\caption{ The multi-step flu outbreak prediction considering the spatio-temporal features. The table presents the MAPEs of RF, SVM, and LSTM models with and without other countries’ flu data. }
			\begin{tabular}{|c|c|c|c|c|c|c|c|c|}
				\hline
				\multirow{2}{*}{Hemisphere}  & \multirow{2}{*}{Country}   & \multirow{2}{*}{step ahead}    &\multicolumn{2}{c|}{SVM} & \multicolumn{2}{c|}{RF} &\multicolumn{2}{c|}{LSTM} \\ \cline{4-9}
				
				&       &     & without & without & with & without & with & without \\ \hline
				\multirow{4}{*}{Southern}      & \multirow{4}{*}{Australia} & 1 & 0.31 & 0.32 & 0.38 &0.37 & 0.30 & 0.23 \\
				&  & 2 & 0.41 & 0.41 & 0.39 & 0.39 &0.30 & 0.25\\
				&  & 3 & 0.52 & 0.52 & 0.48 & 0.46 & 0.33 &0.30 \\
				&  & 4 & 0.66 &0.66 & 0.54 & 0.53 & 0.39 &0.32  \\ \hline
				
				\multirow{4}{*}{Southern}      & \multirow{4}{*}{Brazil} & 1 & 0.28 & 0.28 & 0.30 &0.33 & 0.29 & 0.23 \\
				&  & 2 & 0.34 & 0.33 & 0.85 & 0.31 &0.29 & 0.26\\
				&  & 3 & 0.42 & 0.36 & 0.70 & 0.30 & 0.36 &0.31 \\
				&  & 4 & 0.43 &0.42 & 0.90 & 0.32 & 0.43 &0.36 \\ \hline
				
				\multirow{4}{*}{Northern}      & \multirow{4}{*}{China} & 1 & 0.21 & 0.19 & 0.15 &0.20 & 0.13 & 0.13 \\
				&  & 2 & 0.35 &0.35 &0.28 & 0.29 & 0.20 &0.20 \\
				&  & 3 & 0.56 &0.56 & 0.45 & 0.43 & 0.27 & 0.29  \\
				&  & 4 & 0.74 &0.74 &0.55 & 0.51 & 0.36 & 0.37  \\ \hline
				
				\multirow{4}{*}{Northern}      & \multirow{4}{*}{Japan} & 1 & 0.31 & 0.29 & 0.43 &0.41 & 0.31 & 0.28 \\
				&  & 2 & 0.50 &0.48 &0.56 & 0.48 & 0.39 &0.40 \\
				&  & 3 & 0.60 &0.57 & 0.63 & 0.49 & 0.43 & 0.43  \\
				&  & 4 & 0.78 &0.74 &0.70 & 0.66 & 0.44 & 0.54  \\ \hline
				
				\multirow{4}{*}{Northern}      & \multirow{4}{*}{UK} & 1 & 1.28 & 1.31 & 1.46 &1.22 & 0.69 & 0.86 \\
				&  & 2 & 1.71 &1.66 &2.62 & 2.64 & 0.92 &0.95 \\
				&  & 3 &3.10 & 2.89 &3.51 & 3.08 & 1.13 & 1.17   \\
				&  & 4 & 3.67 &3.27 &4.34 & 5.10 & 0.83 & 1.19 \\ \hline
				
				\multirow{4}{*}{Northern}      & \multirow{4}{*}{USA} & 1 & 0.17 & 0.18 & 0.14 &0.14 & 0.18 & 0.15 \\
				&  & 2 & 0.30 &0.29 &0.21 & 0.21 & 0.23 &0.25 \\
				&  & 3 &0.45 & 0.43 &0.27 & 0.24 & 0.24 & 0.29   \\
				&  & 4 & 0.59 &0.59 &0.30 & 0.29 & 0.29 & 0.30 \\ \hline

			\end{tabular}
		\end{table}


		\begin{figure}
		\centering
		\includegraphics[width=0.7\linewidth]{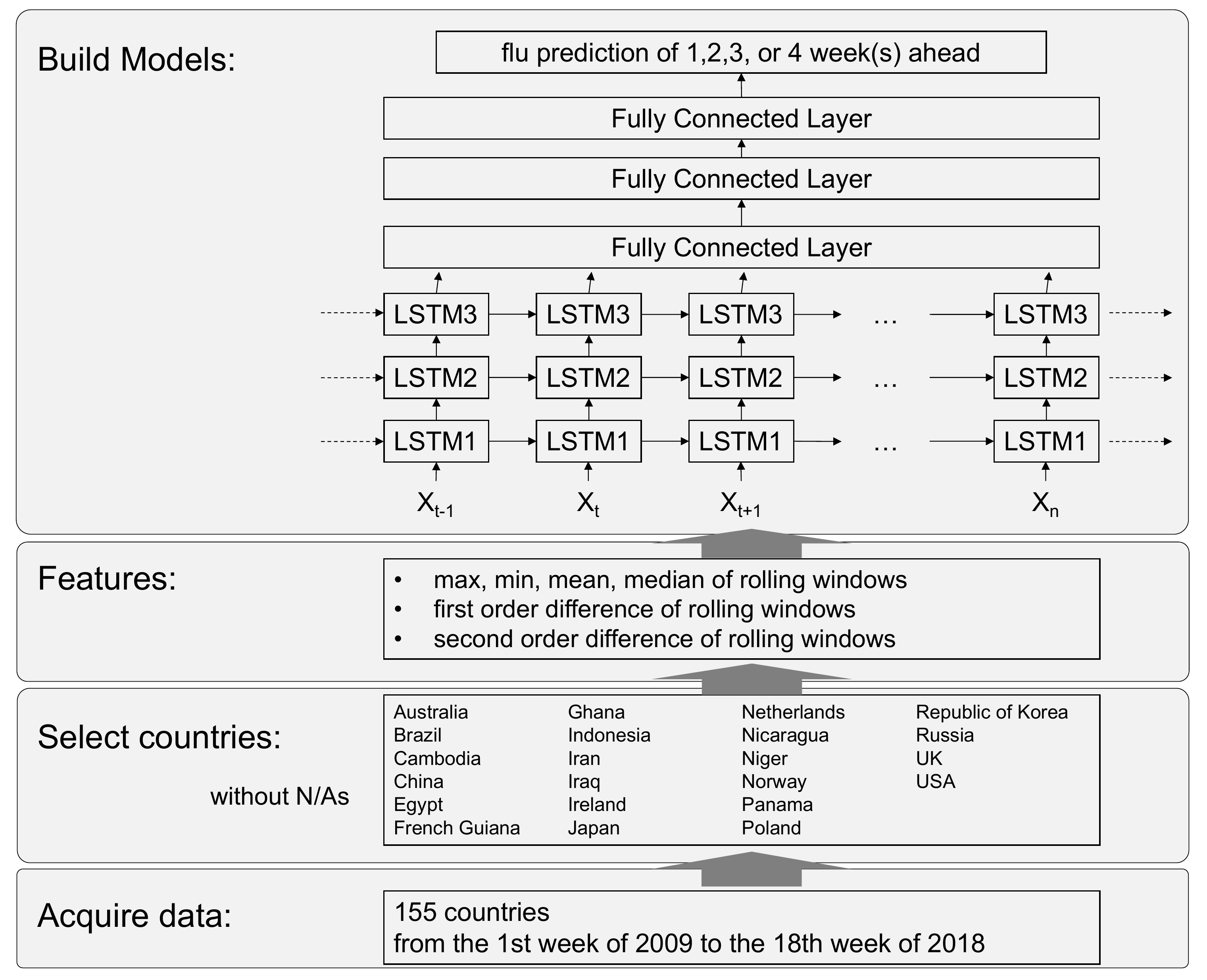}
		\caption{The flow chart of our method.}
		\label{fig:model}
	\end{figure}
	
	\begin{figure}
		\centering
		\includegraphics[width=0.7\linewidth]{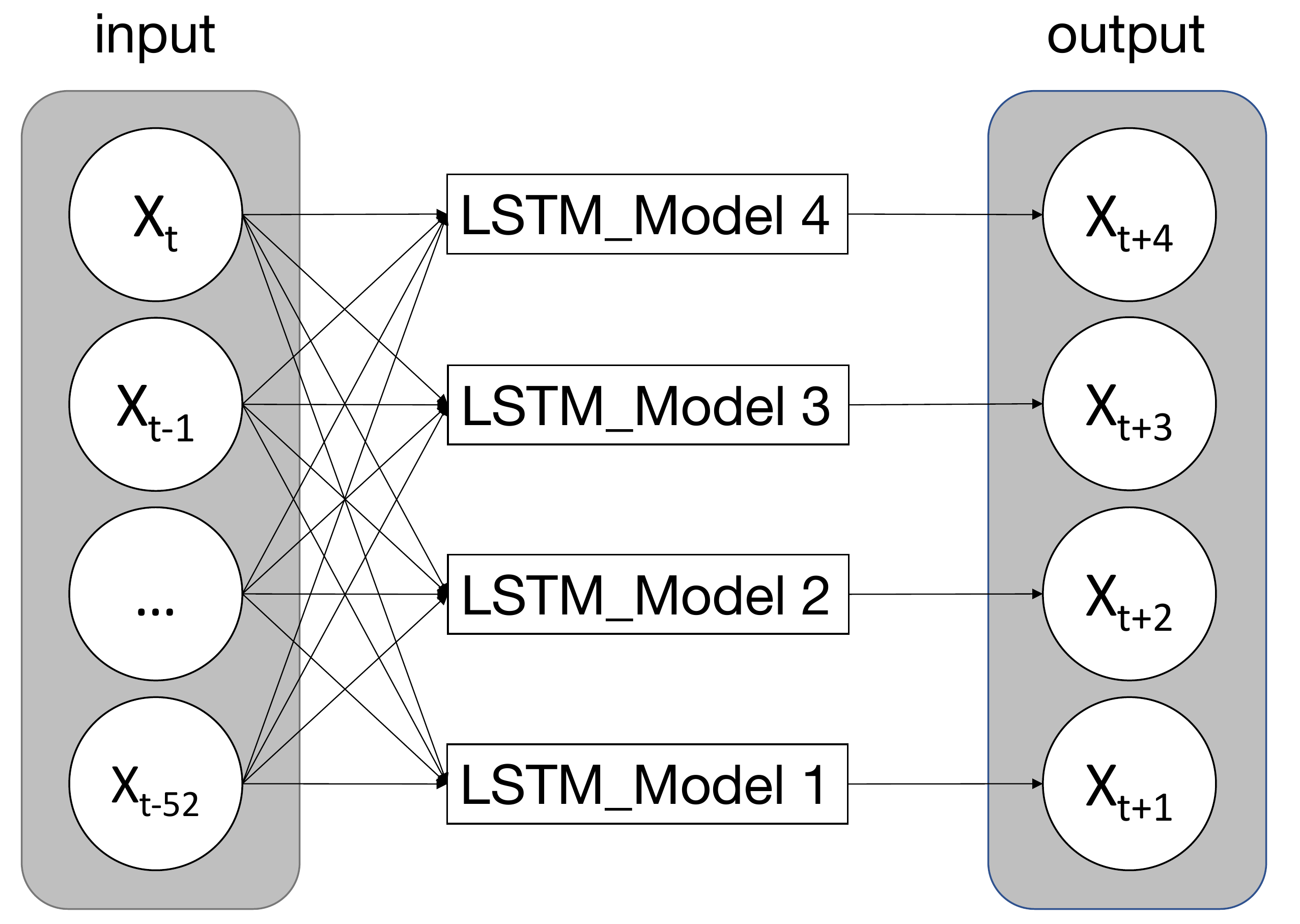}
		\caption{Multi-step prediction.}
		\label{fig:msop}
	\end{figure}

\begin{figure}
	\centering
	\includegraphics[width=0.99\linewidth]{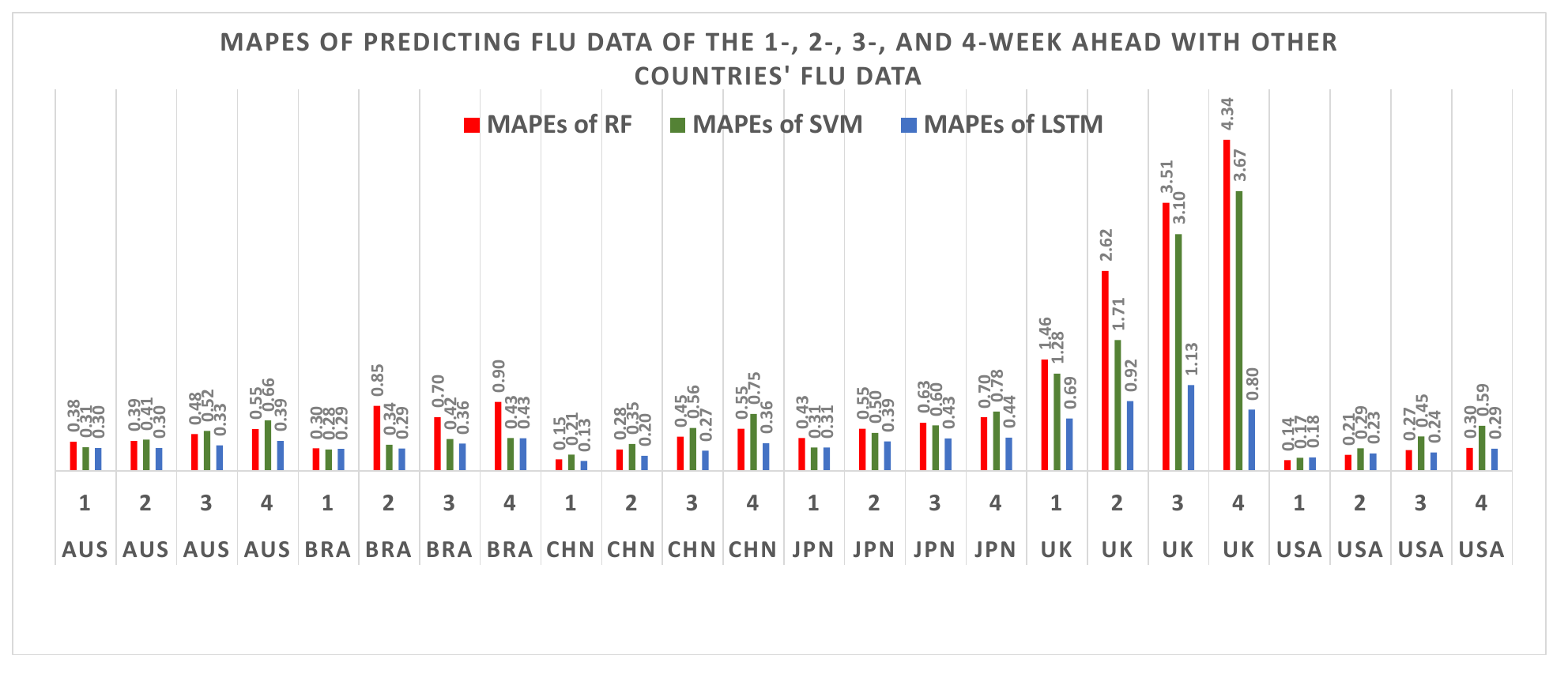}
	\caption{The MAPES considering spatio-temporal features.}
	\label{fig:with}
\end{figure}

	\begin{figure}
		\centering
		\includegraphics[width=0.99\linewidth]{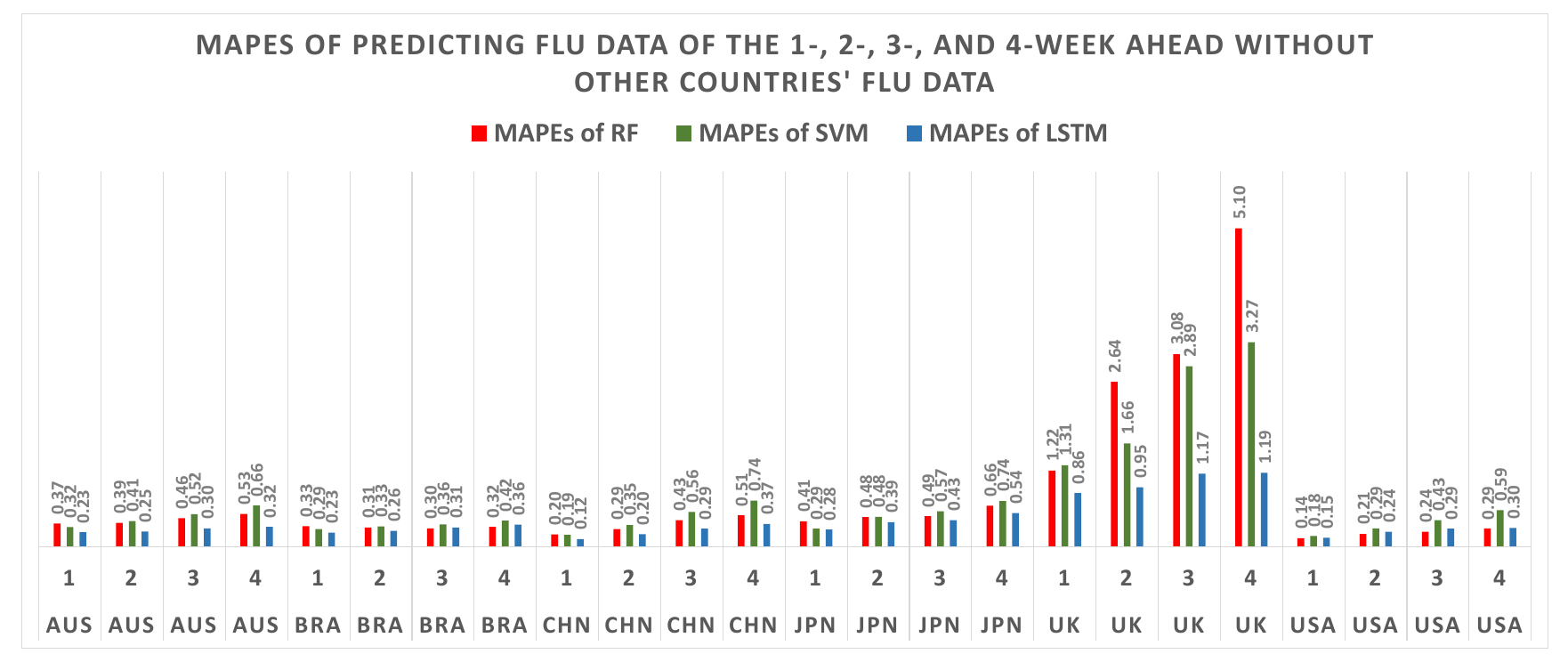}
		\caption{The MAPES without considering spatio-temporal features.}
		\label{fig:without}
	\end{figure}
	
\begin{figure}
	\centering
	\includegraphics[width=0.99\linewidth]{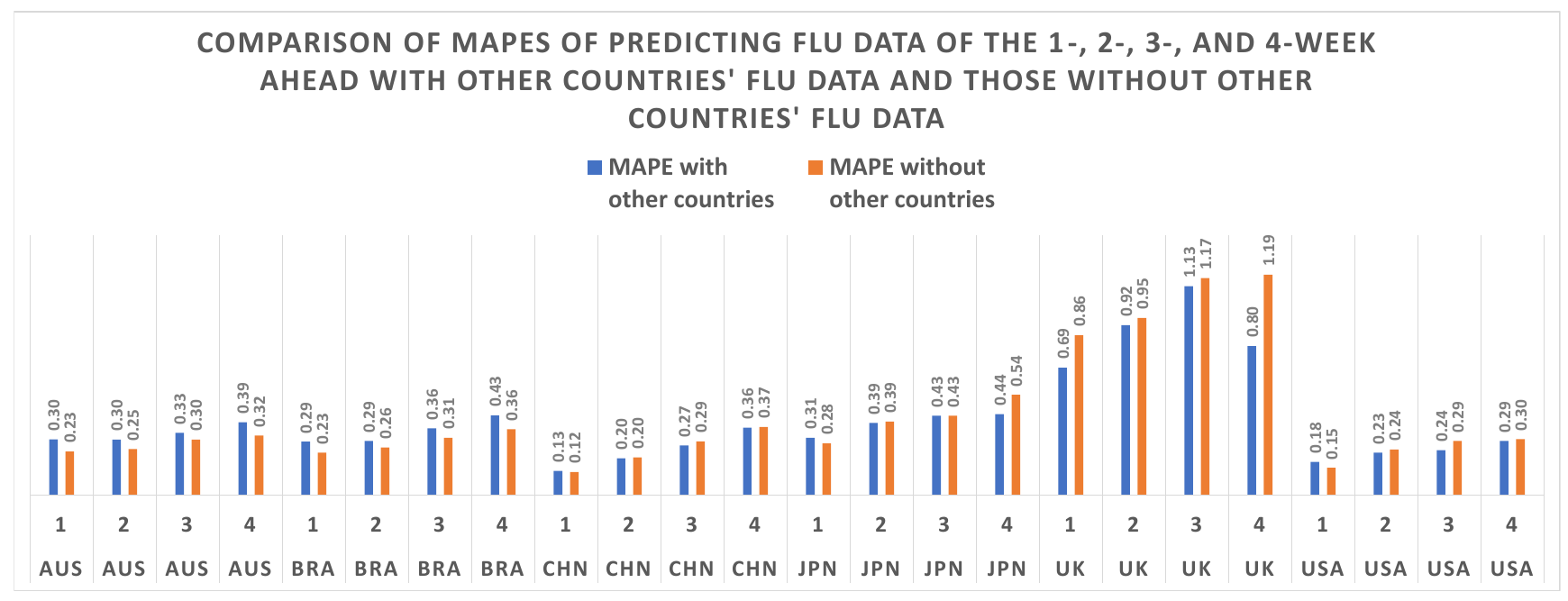}
	\caption{The result of predicting flu outbreaks using LSTM.}
	\label{fig:lstmcompare}
\end{figure}

	\end{backmatter}
\end{document}